# THE DENSITY PROFILE AND THE WAVELENGTH DEPENDENCE OF COMPACT RADIO SOURCE SIZE

Fedor V.Prigara

Institute of Microelectronics and Informatics, Russian Academy of Sciences,

21 Universitetskaya, 150007 Yaroslavl, Russia; fprigara@rambler.ru

## ABSTRACT

In the gaseous disk model, the density profile is derived from the mass and energy conservation in the gravitational field of a central energy source. Using the condition for emission, following from the induced character of radiation process, we obtain the wavelength dependence of radio source size which is consistent with observational data.

*Subject headings: radiation mechanisms: thermal--AGN--pulsars--radio continuum*

## 1. INTRODUCTION

Compact radio sources have the small angular dimensions, usually less than 1 mas, and exhibit the high brightness temperatures. These properties are common for pulsars (Shklovsky 1984), masers (Bochkarev 1992), and active galactic nuclei (Bower & Backer 1998, Kellermann, Vermeulen, Zensus & Cohen 1998). The brightness temperatures of OH masers have the magnitude $T_b \leq 10^{12} K$, and those of water masers have the magnitude $T_b \leq 10^{15} K$ (Bochkarev 1992). Compact extragalactic sources (AGNs) exhibit brightness temperatures in the range of $10^{10}$ K to $10^{12}$ K (Bower & Backer 1998, Kellermann et al. 1998), so these temperatures have an order of magnitude of those of OH masers.

Another feature, which is common for the compact sources, is that their radio emission so far has not received a satisfactory explanation. In particular, it is true for pulsars (Qiao et al. 2000, Vivekanand 2002). In the case of maser sources the modern theory uses some chance coincidences which hardly can maintain in a more profound theory (Bochkarev 1992). At last, it was shown recently that spherical accretion models with the synchrotron mechanism of emission cannot explain the flat or slightly inverted radio spectra of low-luminosity active galactic nuclei (Nagar, Wilson & Falcke 2001, Ulvestad & Ho 2001). The synchrotron self-absorption produces a change in the polarization position angle across the spectral peak. No such a change was detected in gigahertz-peaked spectrum sources (Mutoh et al. 2002).

The difficulties encountered by plasma mechanisms of radio emission from pulsars clearly show that the radiation is produced by a low-energy medium (Gedalin, Gruman & Melrose 2002). Such a medium is the gaseous disk surrounding the central energy source.

## 2. THE GASEOUS DISK MODEL

It was shown recently (Prigara 2001b) that thermal radio emission has a stimulated character. According to this conception thermal radio emission of non-uniform gas is produced by an ensemble of individual emitters. Each of these emitters is a molecular resonator the size of which has an order of magnitude of mean free path $l$ of photons

$$l = \frac{1}{n\sigma} \quad (1)$$

where *n* is the number density of particles and σ is the photoabsorption cross-section.

The emission of each molecular resonator is coherent, with the wavelength

$$\lambda = l, \quad (2)$$

and thermal radio emission of gaseous layer is incoherent sum of radiation produced by individual emitters.

The condition (2) implies that the radiation with the wavelength λ is produced by the gaseous layer with the definite number density of particles *n* .

In the gaseous disk model, describing radio emitting gas nebulae (Prigara 2001a), the number density of particles decreases reciprocally with respect to the distance *r* from the energy center

$$n \propto r^{-1}. \quad (3)$$

Together with the condition of emission (2) the last equation leads to the wavelength dependence of radio source size:

$$r_\lambda \propto \lambda. \quad (4)$$

The relation (4) is indeed observed for sufficiently extended radio sources. For instance, the size of radio core of galaxy M31 is 3.5 arcmin at the frequency 408 MHz and 1 arcmin at the frequency 1407 MHz (Sharov 1982).

### 3. THE WAVELENGTH DEPENDENCE OF RADIO SOURCE SIZE

In the case of compact radio sources instead of the relationship (4) the relationship

$$r_\lambda \propto \lambda^2 \quad (5)$$

is observed (Lo et al. 1993, Lo 1982). This relationship may be explained by the effect of a gravitational field on the motion of gas which changes the equation (3) for the equation

$$n \propto r^{-1/2}. \quad (6)$$

The mass conservation in an outflow or inflow of gas gives *nvr=const,* where v is the velocity of flow. In the gravitational field of a central energy source the energy conservation gives

$$v = (v_0^2 + c^2 r_s / r)^{1/2} \quad (7)$$

where $r_s$ is the Schwarzschild radius. Therefore, at small values of the radius the equation (6) is valid, whereas at the larger radii we obtain the equation (3).

It is well known (Shklovsky 1984) that the delay of radio pulses from pulsars at low frequencies is proportional to $\lambda^2$. This fact is a mere consequence of Eq.(5), if we only assume the existence of the radial density wave traveling across the radius with a constant velocity and triggering the pulse radio emission. In this treatment the pulsars also obey the $\lambda^2$ dependence of compact source size. Note that the wavelength dependence of a pulse duration is a similar effect.

The spatial distribution of SiO, water, and OH masers (each of which emits in own wavelength) in the maser complexes also is consistent with the $\lambda^2$ dependence of compact source size (Bochkarev 1992, Eisner et al. 2001).

To summarize, extended radio sources are characterized by the relation (4), and compact radio sources obey the relation (5).

Thermal radiation in a magnetic field is polarized (see, e.g., Lang 1974). Together with the magnetic field profiles this produces the wavelength dependence of polarization in the form of $p \propto \lambda^{-1}$ for extended sources and $p \propto \lambda^{-2}$ for compact radio sources (Prigara 2002).

## 4. CONCLUSIONS

The compact radio sources are characterized by the following properties: 1) the small angular dimensions; 2) the high brightness temperatures; 3) the $\lambda^2$ dependence of radio source size; 4) the maser mechanism of radio emission.

The properties of the compact radio sources and, in particular, the wavelength dependence of radio source size can be adequately described within the gaseous disk model.

The author is grateful to the anonymous referee for useful comments.